# Measurement of the $\beta$ parameter of activated $CaCO_3$ using time-resolved luminescence spectroscopy

Stephen Tapsak[1], Brielle Hunt[1], John H. Huckans[2]

[1]Department of Environmental, Geographical and Geological Sciences

[2]Department of Biochemistry, Chemistry, Engineering and Physics

Commonwealth University of Pennsylvania, Bloomsburg Campus

Bloomsburg, PA 17815

## Abstract

The decay of luminescence emitted by an activated natural calcite sample after being excited by longwave ultraviolet light (370 nm) has been measured and analyzed. The time evolution of the light intensity did not follow a single exponential decay. Rather, a distribution of decay-times was inferred via the extraction of a fit parameter characterizing the nature of the observed "stretched" exponential decay, herein referred to as the $\beta$ parameter. In conjunction with the average wavelength of the emitted light as well as its average decay-time, the $\beta$ parameter may serve as an important predictor of the nature, concentration and spatial homogeneity of the activators (and possible quenchers) within the calcite sample.

## Introduction

The long-term decay of light emitted by minerals and other materials exposed to high-energy photons (or phonons, etc.) is caused by metastable electron "centers" or "traps" in the material, caused by the presence of intentionally inserted or naturally occurring impurities, also known as *activators* [2,3,9]. Well-known examples are man-made objects which glow as well as the so-called "color centers" which occur naturally in diamonds. Calcite or calcium carbonate ($CaCO_3$) consists of a rhombohedral cell consisting of planar $CO_3$ anion groups, and a Ca ion at the center of an equilateral triangle of oxygen atoms. The structure of calcite is similar to that of NaCl with a shortened trigonal axis. Calcite is a well-studied luminescent mineral, with several well understood luminescence centers including $Mn^{2+}$, $Pb^{2+}$, $Ce^{3+}$, $Dy^{3+}$, $Eu^{3+}$ and several others [4].

Because metastable transitions are considered quantum mechanically forbidden, this *luminescence* often occurs on timescales of milliseconds or even seconds. If all metastable centers within a material possessed identical decay rates, then a simple model would predict light intensity decay following a simple (single) exponential. Instead, however, there is often a distribution of decay-times for the ensemble of traps [9,10]. This decay-time distribution may cause a striking effect in which a single crystal's color changes as it luminesces.

More than a curiosity however, this multi-decay-time (and occasional color-shifting) response, when carefully analyzed can provide important information about the nature of the distribution of the activators (and possible quenchers) within a mineral. Specifically, one fit parameter in particular, the $\beta$ parameter may provide important insight into the types and concentrations of the activators, their coordination, etc. We should note that our investigation is certainly not the first time-resolved spectroscopic study of mineral luminescence [2,4,9,10]. In fact, prior investigations have extracted these important fit parameters (including the $\beta$ parameter) specifically from natural as well as synthetic calcite [5].

Here, we apply the appropriate mathematical model to characterize the decay of luminescence from a natural $CaCO_3$ sample at 300 K following excitation by ultraviolet (UV) light at 370 nm. We extract the average decay-time $\tau_0$ and $\beta$ parameter with high precision and then use these fit parameters to calculate a distribution of decay channels through the activators. The luminescence wavelength and character of the calculated decay channel distribution (most importantly its width), gives support to the hypothesis that this natural calcite sample is activated by spatially inhomogeneous $Mn^{2+}$ substitutions and also possess natural quenchers, such as $Fe^{2+}$.

## Methods

The first quantitative studies of the time evolution of luminescence were carried out by Edmond Becquerel (1820–1891) and published in 1861 [1]. Edmond Becquerel, father of the more famous Henri Becquerel, experimented with various fitting functions for his measured light decay data including simple exponentials as well as various sums of two such exponentials [2]. He also concluded that for certain inorganic solids, an empirical decay function of the form

$$I(t) = \frac{1}{(1+at)^2} \tag{1}$$

gave better fits than a sum of two exponentials. He later proposed a more general function,

$$I(t) = \frac{1}{(1+at)^p} \tag{2}$$

with $p$ taking values between 1 and 2 [2]. This function decays faster than a hyperbola (for which $p = 1$) and is yet slower than a simple exponential. It is similar to a type of decay known as a Kohlrausch function or "stretched exponential." Equation 2 is often referred to as the Becquerel decay law [3].

The light intensity as a function of time, $I(t)$, emitted by a luminescent material is related to the population of excited electrons, $N_e$, as follows

$$I(t) = \frac{\mathrm{d}N_e(t)}{\mathrm{d}t}. \tag{3}$$

As discussed, the phenomenon of luminescence is due to the presence of metastable states in a material interposed between ground and excited electronic states through which excited electrons pass on their way to the ground state. The diversity of these decay channels causes the master equation governing the evolution of the excited electrons to be exceedingly complicated. However, a phenomenological rate equation proposed in [1] based on simple exponential decay may be expressed as a potentially non-linear differential equation as follows,

$$\frac{\mathrm{d}N_e}{\mathrm{d}t} = -kN_e^{\,2-\beta} \tag{4}$$

where $0 \leq \beta \leq 1$ and $k$ is the rate constant. The values $\beta = 1$ and $\beta = 0$ correspond to simple exponential and hyperbolic (for which $p = 1$ in Eq. 2) evolution, respectively. Integration of Eq. 4 leads to

$$N_e(t) = \frac{1}{\left[1+(1-\beta)\frac{t}{\tau_0}\right]^{1/(1-\beta)}} \tag{5}$$

where $\tau_0$ is the decay-time associated with the rate constant $k$. Equation 5 can be thought of as a more general description of Becquerel decay (Eq. 2).

It is also possible to recast Eq. 4 as a linear differential equation with a time-dependent rate constant:

$$\frac{\mathrm{d}N_e}{\mathrm{d}t} = -k(t)N_e \tag{6}$$

$$k(t) = \frac{1}{\tau_0 + (1-\beta)t} \tag{7}$$

where $\tau_0$ is now associated with the *average* decay rate, $\langle k \rangle$. Interestingly, the solutions to Eqns. 4 and 6 (with k defined as above) both yield Eq. 5, from which we may extract the $\beta$ parameter of our sample. We may also model the luminescent decay of our sample as the result of a probability distribution $H(k)$ of decay channels, parametrized by the decay rate $k$.

$$N_e(t) = \int_0^\infty H(k)e^{-kt}\,\mathrm{d}k. \tag{8}$$

Taking the inverse Laplace transform of Eq. 8 to determine $H(k)$, we obtain

$$H(k) = \frac{\left(\frac{1}{(1-\beta)}\frac{k}{\langle k \rangle}\right)^{\beta/(1-\beta)} \exp\left(-\frac{1}{(1-\beta)}\frac{k}{\langle k \rangle}\right)}{(1-\beta)\langle k \rangle \Gamma\left(\frac{1}{1-\beta}\right)} \tag{9}$$

with an average decay rate (of the ensemble) denoted $\langle k \rangle$. After extracting the $\beta$ parameter and $\tau_0$ from a fit (Eq. 5) to the time-resolved luminescence data using, we then employ Eq. 9 (where $\langle k \rangle$ is taken to be the inverse of the $\tau_0$ fit parameter) to calculate the probability distribution of decay channels associated with the activators within the calcite sample.

All luminescence spectra and time-resolved data were generated using a uvBeast V3 light source which was extinguishable in less than 1 ms. The source spectrum centered at approximately 370 nm is shown on Fig. 1. This UV source was focused with a 51-mm, 50-mm focal length biconvex lens (Ealing Optics #A23-8907) down to a 5-mm spot size on the surface of a 1.0 mm thick sliced natural calcite sample behind which the luminescence was captured using an Ocean Optics TP300-UV-VIS optical fiber connected to a USB4000 spectrometer controlled by OceanView spectroscopy software (v.1.6.7). See Fig 2. All luminescence spectra and decay-times were measured at 300 K. The natural calcite sample originated from Bloomsburg, PA.

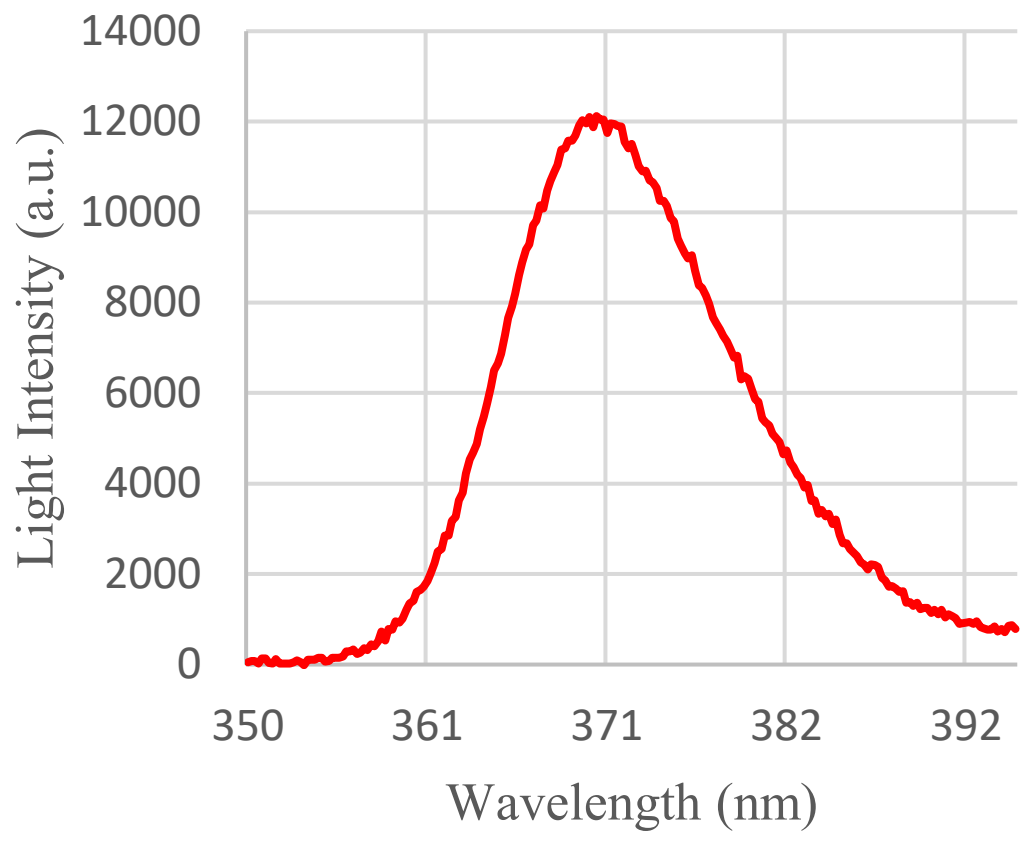

FIG. 1: Ultraviolet source spectrum used to illuminate natural calcite sample.

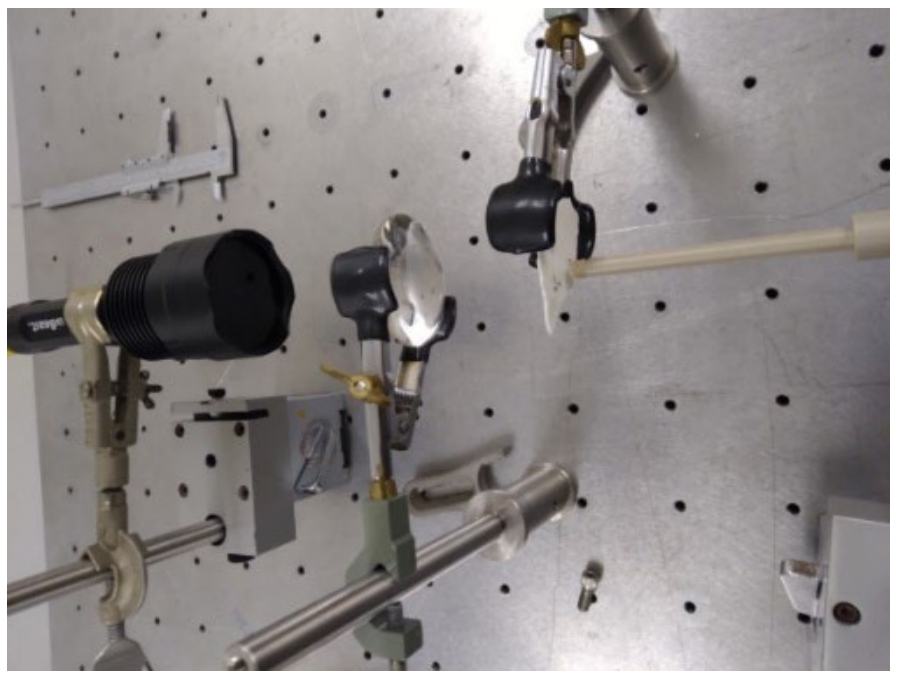

FIG. 2: Experimental setup with source (left) focused onto sample (center) with luminescence captured by fiber (right).

## Results and Discussion

Shown on Fig. 3 is the spectrum of luminescence emanating from the activated $CaCO_3$ sample following excitation by the 370 nm source light. The spectrum is clearly peaked at approximately 645 nm which supports our assumption that the sample is activated primarily by $Mn^{2+}$ [7,8]. Figure 4 depicts the time-resolved decay of the 645 nm luminescence emitted by the calcite sample. The obvious curvature of the fitted trendline in Fig. 4 (in red) to the decay data when plotted on a semi-log plot is a clear signature of the non-single exponential decay of the 645 nm luminescence. The average decay-time, $\tau_0$, and $\beta$ parameter obtained were 34(2) ms and 0.78(5), respectively, with less than 10% uncertainty in both values.

On the basis of these fit parameters, and using Eq. 9, we have calculated the distribution of decay channels associated with the activators within the $CaCO_3$. As shown on Fig. 5, the calculated probability distribution corresponds to a Gaussian which is positively skewed toward faster decay rates. The average decay rate of 29 $s^{-1}$ (inverse of the 34 ms average decay-time obtained from the fit) is approximately 20% higher than the most probable decay rate of 23 $s^{-1}$. Furthermore, the distribution is rather broad with approximately an order of magnitude difference between the slowest (roughly 5 $s^{-1}$) and the fastest (70 $s^{-1}$) decay rates. This information in particular concerning the broadness of the distribution would seem to suggest a spatially inhomogeneous concentration of $Mn^{+2}$ substitutions within the natural calcite sample, or perhaps irregularly interspersed quenchers such as $Fe^{2+}$, iron being the most common and important quencher in natural calcite [6]. Due to the fact that there was only one luminescence color (645 nm) and no color-shifting, it is unlikely that our sample possessed activators besides the assumed dominant $Mn^{2+}$ in this sample. However, it is possible that by using a grating spectrometer with better spectral resolution, the peak observed in Fig. 3 would have been shown to actually be composed of several narrower peaks, perhaps not all due to $Mn^{2+}$ activators.

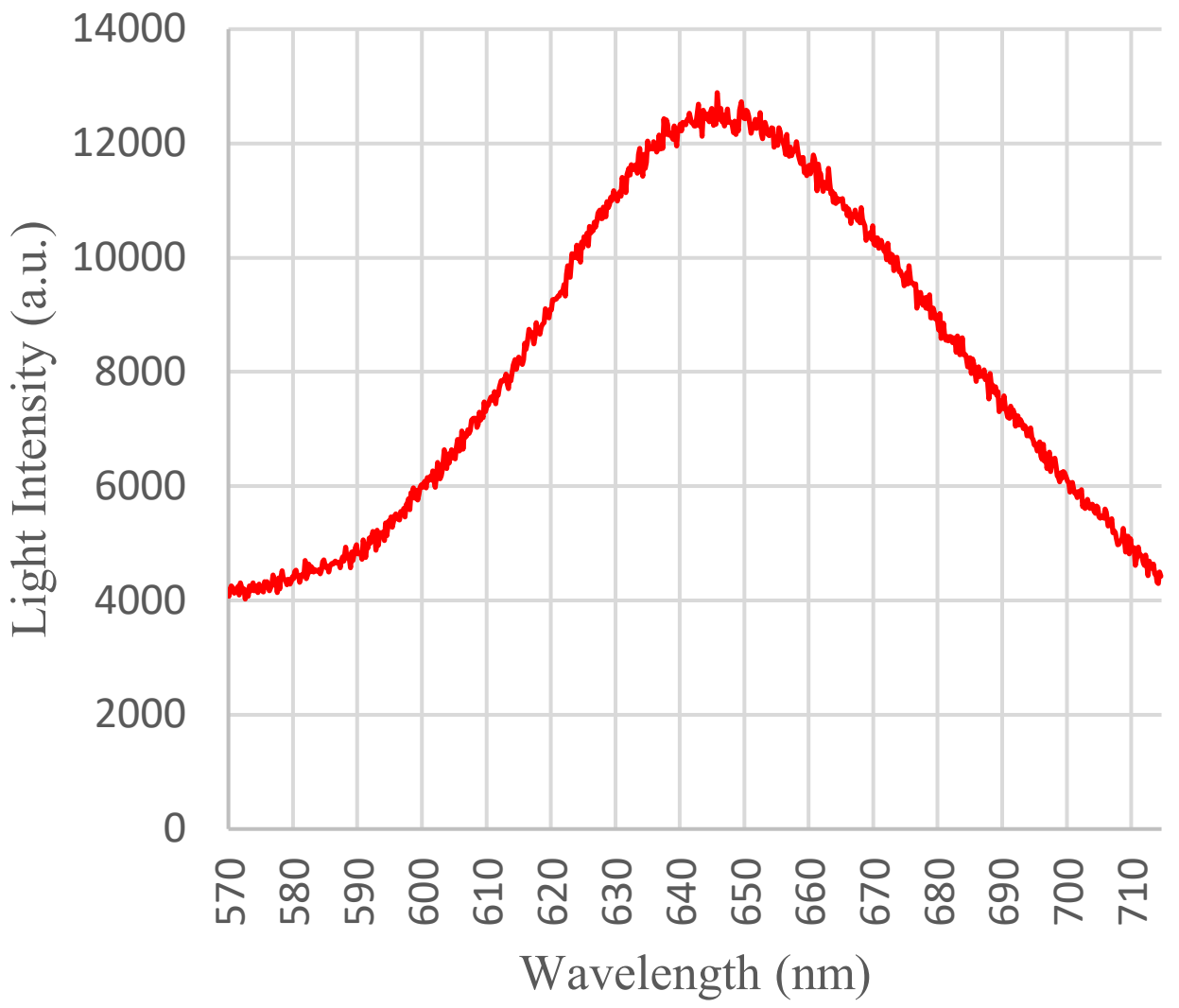

FIG. 3: Red luminescence spectrum of natural calcite sample following longwave UV excitation. The peak is centered at 645 nm with a FWHM of approximately 90 nm.

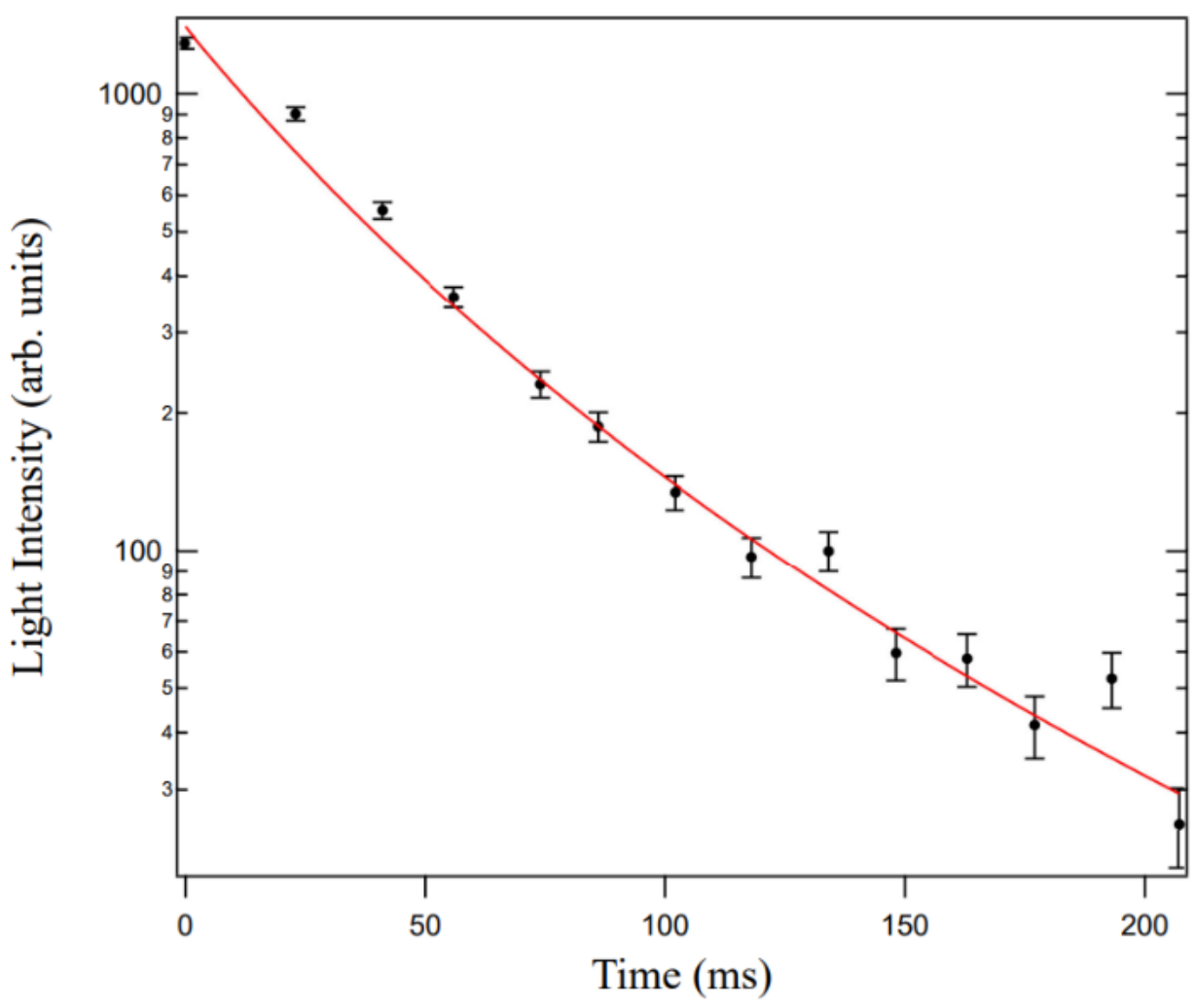

FIG. 4: Time-resolved decay of 645 nm luminescence from natural calcite sample. The average decay-time $\tau_0$ and $\beta$ parameter obtained from the fit (in red) are 34(2) ms and 0.78(5), respectively.

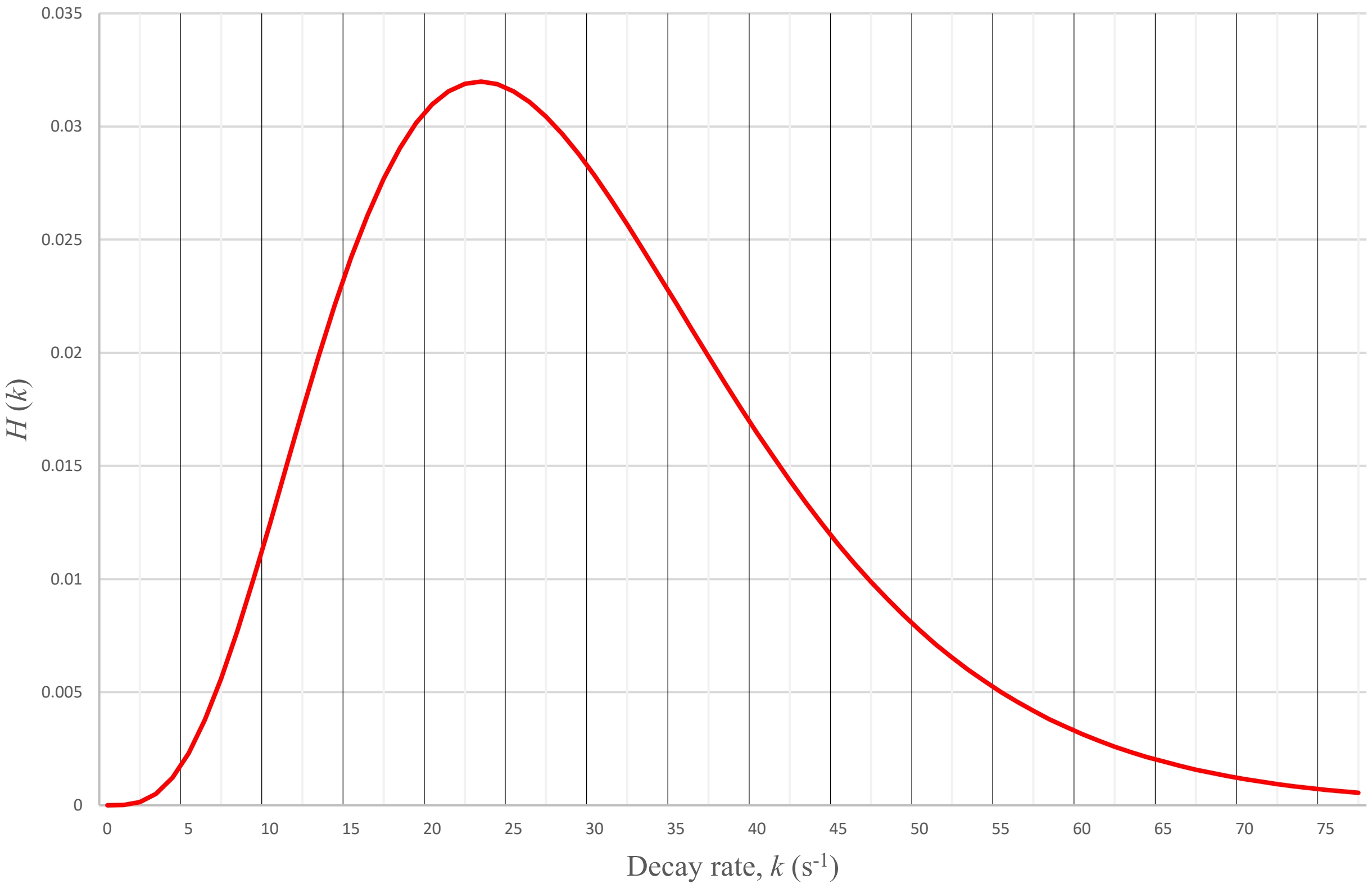

FIG. 5: Calculated probability distribution of luminescence decay channels in a natural calcite mineral sample, exhibiting positive skew. The mode of the distribution occurs at $k = 23\ s^{-1}$. The average decay rate $\langle k \rangle$ of the distribution is $29\ s^{-1}$.

## Conclusion

The decay of luminescence emitted by an activated natural calcite sample after being excited by longwave ultraviolet light was measured and analyzed. In addition to a determination of the spectrum of emitted luminescence centered at 645 nm, a time-resolved spectroscopic analysis allowed for the extraction of the average decay-time $\tau_0$ as well as the $\beta$ parameter characterizing the stretched decay of the luminescence. Taken together, these data enabled us to make reasonable predictions about the nature, concentration, spatial homogeneity, etc. of the activators (and possible quenchers) within the calcite sample. In the future, we intend to test these predictions using refined time-resolved techniques involving spatial scans of the calcite sample with a more tightly focused UV source or perhaps complementary techniques such as cathodoluminescence (CL) microscopy or X-ray fluorescence (XRF).

Acknowledgments

We wish to acknowledge helpful discussions with Dr. Adrian Van Rythoven of the Montana Bureau of Mine and Geology as well as the technical assistance of Dr. Stephen Whisner of the Bloomsburg campus of Commonwealth University of Pennsylvania in the preparation of the calcite samples used in our study.